# Characterization of 0.18μm CMOS Ring Oscillator at Liquid Helium Temperature

Chao Luo, Tengteng Lu, Zhen Li, Jie He, Guoping Guo
Key Laboratory of Quantum Information
University of Science and Technology of China
Hefei, China
gpguo@ ustc.edu.cn

Chao Luo, Tengteng Lu, Zhen Li
Department of physics
University of Science and Technology of China
Hefei, China
lc0121@mail.ustc.edu.cn

*Abstract*—**This paper presents low power dissipation, low phase noise ring oscillators (ROs) based on Semiconductor Manufacturing International Corporation (SMIC) 0.18μm CMOS technology at liquid helium temperature (LHT). First, the characterization and modelling of CMOS at LHT are presented. The temperature-dependent device parameters are revised and the model then shows good agreement with the measurement results. The ring oscillator is then designed with energy efficiency optimization by application of forward body biasing (FBB). FBB is proposed to compensate for the threshold voltage ($V_{TH}$) shift to preserve the benefits of the enhancement of the carrier mobility at 4.2K. The delay per stage ($\tau_P$), the static current ($I_{STAT}$), the dynamic current ($I_{DYN}$), the power dissipation (P) and the phase noise ($L(f_{off})$) are analyzed at both 298 K and 4.2 K, with and without FBB. The performance of the designed RO in terms of speed ($\tau_P$=179ps), static current (23.55nA/stage), power dissipation (2.13μW) and phase noise (-177.57dBc/Hz@1MHz) can be achieved at 4.2K with the supply voltage ($V_{DD}$) reduced to 0.9V.**

## I. INTRODUCTION

With the development of quantum computing, the wiring requirements between the cryogenic quantum processor and the room temperature (RT) read-out controller can be both expensive and unreliable [1]. As an alternative, a cryogenic electronic interface that consists of control, interaction and read-out stages has been proposed. Cryogenic CMOS technology (cryo-CMOS) offers a scalable solution for quantum device interface fabrication [2]. The characteristics of metal-oxide semiconductor field-effect transistors (MOSFETs) change because of freeze-out effect, thus stimulating a requirement for a revised SPICE model of these devices and circuit designs to be developed for use at cryogenic temperatures.

Therefore, this paper describes a study of the characterization of SMIC 0.18μm CMOS ROs down to 4.2K. Characterization and modelling of MOSFETs are performed. The RO performance in terms of delay per stage ($\tau_P$), static current ($I_{STAT}$), dynamic current ($I_{DYN}$), power dissipation (P) and phase noise ($L(f_{off})$) is studied at both 298K and 4.2K for $V_{DD}$ ranging from 0.9V to 1.8V. Cryogenic effects on the RO performance are studied with and without FBB. The ROs have lower phase noise at 4.2K than at room temperature, while the phase noise would slightly increase as the FBB is applied. It is demonstrated that $V_{DD}$ can be reduced down to 0.9V by appropriate application of body biasing while maintaining an ultralow total power dissipation. In addition, the application of FBB takes advantage of both equivalent speed and lower power dissipation at the low-$V_{DD}$ case, compared to the case of high-$V_{DD}$.

TABLE I. SUMMARY OF CHARACTERIZED DEVICES

| Device Size | Thin-oxide (3.87nm) NMOS/PMOS | |
|---|---|---|
|  | *W(μm)* | *L(μm)* |
| SMIC 0.18 μm technology | 0.22 | 0.18 |
|  | 10 | 10 |
|  | 10 | 0.6 |
|  | 10 | 0.2 |
|  | 10 | 0.18 |
|  | 10 | 0.16 |

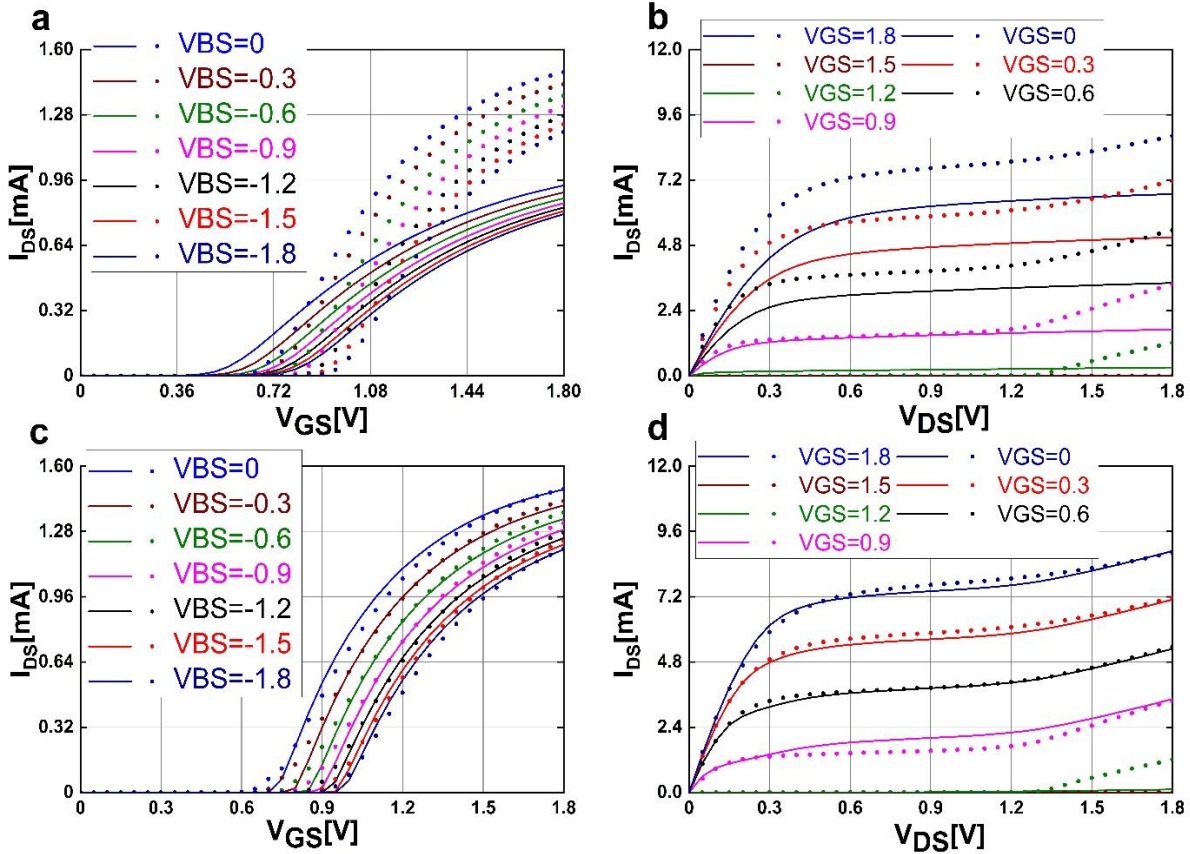

Figure 1 Device measurement and modelling, device size W/L=10μm /0.16μm. (a,b) $I_{DS}$-$V_{GS}$ curves (a) and $I_{DS}$-$V_{DS}$ (b) at RT (solid lines) and LHT (dashed lines); (c,d) Modelling (solid lines) and measurements (dashed lines) of at LHT

## II. CMOS DEVICE MODELLING

The measurements of the CMOS transistors were performed using thin-oxide (3.87 nm) SMIC 0.18μm technology and for a wide range of feature sizes (TABEL I). All electrical measurements were performed using an Agilent B1500A semiconductor device analyzer. The cryogenic measurements were performed in the liquid helium Dewar. The most noticeable irregularity in the drain-source current-voltage ($I_{DS}$-$V_{DS}$) characteristics (Fig.1b) is the kink that occurs in the mid- $V_{DS}$ range at 4.2K. This phenomenon is ascribed to the self-polarization of the bulk at cryogenic temperatures [3]. In this paper, temperature-dependent parameter extraction and a sub-circuit model were applied to eliminate the deviations between the RT BSIM3 model and the measured values [4,5]. The extraction procedure was performed using BsimProPlus and model revision was performed using Matlab. As shown in Fig.1, good agreement with the measurements was achieved for devices operating at LHT. Significant optimization of the BSIM3 model thus been achieved.

## III. CIRCUIT DESCRIPTION

The gate width/length of both NMOS and PMOS transistors are 0.22μm/0.18μm. The layout of RO is shown schematically in Fig. 2a, while Fig. 2b shows a schematic of the single inverter stage composed of NMOS and PMOS transistors. The RO consists of 101 identical stages combined with an enabling two-way AND gate. The parameters and measurements protocols are listed in Table II.

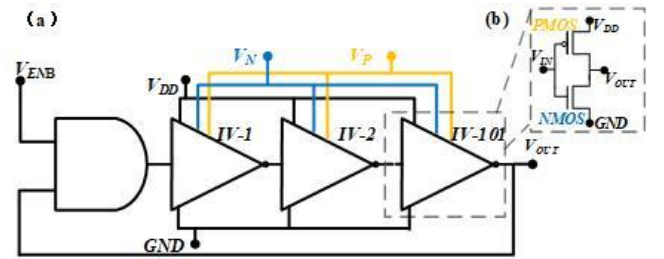

Figure 2 Schematic of 101-stages RO circuit (a) and single inverter stage (b).

The impulse sensitivity function (ISF) is introduced to describe the sensitivity of every point of the wave form to a perturbation in single-ended ring oscillators [6] and the single-

sideband phase-noise spectrum due to a white-noise current source is given by [7]

$$L\{f_{off}\} = \frac{\Gamma_{rms}^2}{8\pi^2 f_{off}^2} \cdot \frac{i_n^2/\Delta f}{q_{max}^2} \quad (1)$$

where $\Gamma_{rms}$ is the root mean square (RMS) value of the ISF, $i_n^2/\Delta f$ the single-sideband power spectral density of the noise current source, $q_{max}$ given by the relation $q_{max} = C_L V_{DD}$, $f_{off}$ the frequency offset from the carrier.

According to [6], the approximate expression for $\Gamma_{rms}$ is obtained:

$$\Gamma_{rms} = \sqrt{\frac{2\pi^2}{3\eta^3}} \frac{1}{N^{1.5}} \quad (2)$$

where $\eta$ is a proportionality constant, which is typically close to one and N is the number of stages of RO.

For single-ended N-stages CMOS ring oscillators, assuming that the thermal noise sources of every inverter are uncorrelated and that the waveform (hence the ISF) of all nodes are the same except for a phase-shift, the total phase noise is N times the value given by (1). Taking only these inevitable noise sources into account, the expression for phase noise is given by [6]:

$$L\{f_{off}\} \approx \frac{8}{3\eta} \cdot \frac{\kappa T}{P} \frac{V_{DD}}{V_{char}} \cdot \frac{f_0^2}{f_{off}^2} \quad (3)$$

where $\kappa$ is the Boltzmann constant, $T$ the temperature, $V_{DD}$ the operating voltage, $f_0$ the frequency of RO, $P$ the total power dissipation, $V_{char}$ the characteristic voltage of the device.

The total power dissipation $P$ is approximately given by
$$P = 2\eta N V_{DD} q_{max} f_0 \quad (4)$$

For short-channel mode of operation, the $V_{char}$ is defined as

$$V_{char} = \frac{E_c L}{\gamma} \quad (5)$$

where $E_c$ is defined as the value of electric field resulting in half the carrier velocity expected from low field mobility, $L$ is the gate-length and $\gamma$ is 2/3 for long-channel devices in saturation region and 2~3 times greater for the case of short-channel devices. Note the absence of any dependency on the number of stages in (3).

TABLE II. RO PARAMETERS, UNITS, DESCRIPTION, AND THE MEASUREMENT PROTOCOL

| Parameter | Unit | Description | Measurement protocol |
|---|---|---|---|
| $f$ | Hz | RO frequency | $V_{ENB}=V_{DD}$, $V_{CC}=1V$ and measure OUT. |
| $\tau_p$ | ps | RO delay per stage | $=1/(f \times 2 \times N)$ |
| $I_{STAT}$ | pA/stage | Static current per stage in non-oscillating state | $V_{ENB}= V_{CC}$=GND and measure OUT. |
| $I_{DYN}$ | nA/stage | Dynamic current per stage in oscillating state | $V_{ENB}=V_{DD}$, $V_{CC}$=GND and measure OUT |
| $P$ | μW | Total Power dissipation | $=(I_{DYN}- I_{STAT}) \times V_{DD}$ |
| $L(f_{off})$ | dBc/Hz | Phase noise | Equation (3) |

## IV. SIMULATION RESULTS

The static current per stage generated by leakage from the supply to ground of the 101-stages RO is shown in Table 2 for a set of $V_{DD}$. The static current (and thus the static power consumption) shows a significant decrease with temperature decreases. The delay per stage of CMOS inverter is given approximately by $\tau_p=C_L \times V_{DD}/I_{EFF}$, where $I_{EFF}$ is the effective current [8] and $C_L$ is the load capacitance, which includes the inversion capacitance, the parasitic capacitance and the wiring capacitance between the two stages. Reducing power dissipation by reducing the value of $V_{DD}$ degrades the speed of RO, as shown in Fig. 3.

The mobility of transistor is strongly enhanced because of the reduction in carrier scattering due to lattice vibrations [4] and should thus lead to a smaller τp at LHT for a given $V_{DD}$. However, the RO suffers a significant reduction in operating speed, as shown in Fig.3a. This increase in the delay per stage is ascribed to the $V_{TH}$ shift (Fig. 1a) that occurs at LHT for both NMOS and PMOS transistors. It should be noted that the load capacitance is weakly dependent on temperature and is a minor reason for the speed change of RO [8]. Body biasing was used to compensate the $V_{TH}$ shift to preserve the benefit of higher carrier mobility. On reduction of temperature from RT to 4.2K, the delay per stage shows significant reductions by 75.68% at $V_{DD}$=0.9V, 66.33% at $V_{DD}$=1.2V, 56.04% at $V_{DD}$=1.5V, 48.49% at $V_{DD}$=1.8V. The $\tau_p$ of RO is 100.66ps@$V_{DD}$=1.3V with FBB ($V_N/V_P$=1.1V/-0.6V) and is 100.11ps@$V_{DD}$=1.5V without FBB, while the total power dissipation P=6.96μW and 9.68μW respectively. Equivalent speed is achieved by applying FBB while the power consumption is reduced. Note that the result can be optimized by adjusting the values of FBB. Application of body biasing thus results in significant $\tau_p$ gain especially at low supply voltages, but cause a slight increase in the power dissipation, as shown in Fig. 3 and Fig. 4.

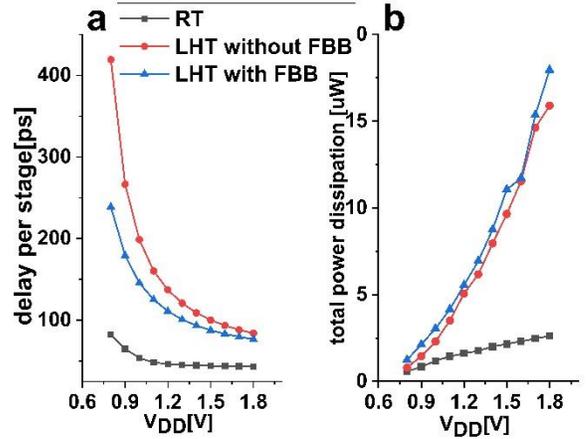

Figure 3. Comparison of ring oscillator performance at room temperature and 4.2K with and without body biasing ($V_N/V_P$=1.1V/-0.6V): (a) comparison of energy delay per stage; (b) comparison of total power dissipation.

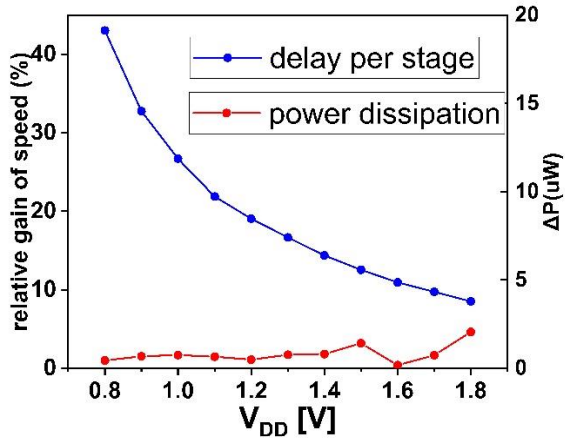

Figure 4. Power dissipation increase and relative enhancement of $\tau_p$

Fig. 4 shows the relative enhancement of $\tau_p$ and the increased total power dissipation with and without FBB for a set of values. The RO performance in terms of $\tau_p$, $I_{DYN}$, $I_{STAT}$, P and $L(f_{off})$ (at a frequency offset of 1MHz) as measured/calculated at 4.2K and 298K with and without FBB is summarized in Table III. The phase noise decreases approximately 25dBc/Hz at 4.2K when compared to that at RT for different VDD and increases slightly when FBB is applied.

TABLE III. COMPARISON OF RO PERFORMANCE AT 298K AND 4.2K BETWEEN THE CASES WITH AND WITHOUT FBB

| T (K) | $V_{DD}$ (V) | Without FBB / with FBB $V_N/V_P$ (V/V) | $\tau_p$ (ps) | $I_{DYN}$ (nA/stage) / $I_{STAT}$ (fA/stage) | | P (μW) | $L(f_{off})$ (dBc/Hz) |
|---|---|---|---|---|---|---|---|
| 298 | 0.9 | | 64.80 | 9.23 | 2907 | 0.84 | -154.65 |
| | 1.2 | | 46.14 | 13.31 | 40594 | 1.61 | -153.17 |
| | 1.5 | | 44.00 | 14.29 | 63208 | 2.15 | -152.97 |
| | 1.8 | | 13.16 | 14.46 | 77089 | 2.61 | -152.88 |
| 4.2 | 0.9 | | 266.44 | 16.04 | 0.10 | 1.46 | -179.30 |
| | | 1.1/-0.6 | 179.17 | 23.55 | 0.16 | 2.14 | -177.57 |
| | 1.2 | | 137.06 | 41.87 | 0.13 | 5.07 | -176.41 |
| | | 1.1/-0.6 | 110.92 | 45.94 | 0.19 | 5.57 | -175.49 |
| | 1.5 | | 100.11 | 63.87 | 0.17 | 9.68 | -175.05 |
| | | 1.1/-0.6 | 87.56 | 73.28 | 0.22 | 11.10 | -174.46 |
| | 1.8 | | 83.79 | 87.66 | 0.20 | 15.94 | -174.27 |
| | | 1.1/-0.6 | 76.66 | 99.02 | 0.26 | 18.00 | -173.89 |

## V. CONCLUSION

This paper describes the electrical characterization of SMIC 0.18μm CMOS ROs at temperatures down to 4.2K. It is demonstrated that by appropriate application of the FBB, benefit in terms of both speed and power dissipation can be derived at low supply voltage of VDD=0.9V. The speed of the designed RO is significantly optimised with a small increase in power consumption. Very small power dissipation of 2.13μW with τp=179ps and phase noise =-177.57dBc/Hz@1MHz for VDD=0.9V are achieved at 4.2K. The results indicate the applicability of body biasing to the design of power-efficient peripheral circuits for large scale quantum computing.

ACKNOWLEDGMENT

The authors would like to thank SMIC for device fabrication and software support. This work was supported by the National Key Research and Development Program of China, China (Grant No.2016YFA0301700), the National Natural Science Foundation of China, China (Grant No.11625419), the Anhui initiative in Quantum information Technologies, China (Grants No.AHY080000). This work was partially carried out at the USTC Centre for Micro- and Nanoscale Research and Fabrication.